\documentclass[a4paper]{jpconf}
\usepackage{graphicx}
\usepackage{siunitx}
\usepackage[symbol]{footmisc}
\usepackage[colorlinks, citecolor = blue, urlcolor = blue]{hyperref}
\renewcommand{\thefootnote}{\fnsymbol{footnote}}

\begin{document}
\title{Characterization of pixelated silicon detectors for daily quality assurance measurements in proton therapy}

\author{Isabelle Schilling$^{1}$, Claus Maximillian Bäcker$^{1,2,3,4}$, Christian Bäumer$^{1,2,3,4,6}$, Carina Behrends$^{1,2,3,4}$, Kevin Kröninger$^{1}$, Beate Timmermann$^{2,3,4,5,6}$, and Jens Weingarten$^{1}$}

\address{$^{1}$TU Dortmund University, Department of Physics, D-44221 Dortmund, Germany}
\address{$^{2}$West German Proton Therapy Centre Essen, D-45122 Essen, Germany}
\address{$^{3}$West German Cancer Center, D-45122 Essen, Germany}
\address{$^{4}$University Hospital Essen, D-45122 Essen, Germany}
\address{$^{5}$University Hospital Essen, Clinic for Particle Therapy, D-45122 Essen,Germany}
\address{$^{6}$German Cancer Consortium, D-69120 Heidelberg, Germany}
\ead{isabelle.schilling@tu-dortmund.de}

\begin{abstract}
The advanced imaging and delivery techniques in proton therapy allow conformal high-dose irradiation of the target volume with high accuracy using pencil beam scanning or beam shaping apertures. These irradiation methods increasingly include small radiation fields with large dose gradients, which require detector systems with high spatial resolution for quality assurance.
In addition the measurement of all success parameters for daily quality assurance with only one proton field and one simple detector system would save a lot of time in clinical usage. Based on their good spatial resolution and high rate compatibility, pixelated silicon detectors could meet the new requirements. To assess their applicability in proton therapy, ATLAS pixelated silicon detectors are used to measure the lateral beam profile with high spatial resolution.
Furthermore, a dose dependent detector calibration is presented to allow the measurement of the requested output constancy.
A strategy to verify the proton energy during the daily quality assurance is under study. Promising results from proof-of-principle measurements at the West German Proton Therapy Centre in Essen, Germany, have been obtained.
\end{abstract}

\section{Introduction}
Proton therapy has been proven to be very effective for the treatment of various kinds of tumors, since it allows to deposit the prescribed dose in a well-defined range \cite{cancer}. 
Pencil beam scanning (PBS) has been quickly becoming the preferred mode of dose delivery. A small diameter beam ($5-10\,\si{\mm}$) is scanned across the target volume in energy layers, ideally irradiating the complete tumor while depositing next to no dose in the surrounding normal tissue laterally and behind the tumor.
This effect can be further increased using apertures, which allow sharper field gradients at the edge of the target volume.

To ensure the safety of the patients and the efficiency of the treatment, Task Group 224, established by the American Association of Physicists in Medicine, has published comprehensive quality assurance (QA) guidelines for proton therapy centers with QA measurements grouped into daily, monthly and annual tasks \cite{aapm}. During the daily quality assurance, amongst other things, the spot shape, output constancy and range of the pencil beam fields has to be verified. This is usually done using different detectors, for example the Lynx PT detector (IBA Dosimetry, Schwarzenbruck, Germany) to characterize the spot shape and the DailyQA3 detector (Sun Nuclear, Melbourne, USA) to verify the output constancy \cite{dailyqa3}. Since currently only few commercially available detector systems allow to perform all measurements with the same detector, setup times for the daily QA measurements are around 30 to 45 minutes.

To address this issue, the feasibility of using silicon pixel detectors, developed for the Insertable B-Layer (IBL) upgrade of the ATLAS Pixel detector at the Large Hadron Collider (LHC) at CERN, is tested. In this paper, we report the results of first proof-of-principle measurements, carried out at clinical beam lines featuring PBS delivery mode at the West German Proton Therapy Centre, Essen (WPE).

\section{Experimental setup}
The ATLAS IBL Pixel detectors are hybrid detectors, with a $\SI{200}{\micro\meter}$ thick n-in-n silicon sensor, segmented in $80\times 336$ pixels with a pitch of $(250\times 50)\,\si{\micro\meter^2}$ \cite{IBL}.
The sensor is bump bonded to a FE-I4B readout chip \cite{readout}. As the detector was designed to track charged particles, individual protons are registered with a hit efficiency in excess of $99\%$. The readout chip provides information on the amount of charge deposited in the sensor, by quantifying the shape of the amplifier output in terms of a discriminator output signal duration (time-over-threshold, ToT). The amplifier gain and the discriminator threshold can be set for every pixel, ensuring uniform response across the detector. These parameters determine the dynamic range for the measurement of the energy deposited in the sensor. The best possible configuration of gain and threshold for the measurements presented here cover the range of about $\SI{100}{\kilo\eV}$ to $\SI{600}{\kilo\eV}$ deposited in the silicon sensor.
Measurements to assess the applicability of the ATLAS IBL detectors for daily QA were carried out at the PBS lines of the WPE using both uncollimated single pencil beams with an expected spot size of $\sigma=\SI{5.53}{\milli\metre}$ as well as scanned homogeneous fields of $(2.5 \times 2.5)\,\si{\cm^2}$. The proton energy is $\SI{100}{\mega\eV}$ unless stated otherwise. The detector was mounted on a goniometer in the isocenter.

\section{Results}
Characterizing measurements should confirm the ability to improve the quality assurance by using pixelated semiconductor detectors. Therefore the data are analyzed regarding the required parameters for the quality assurance in proton therapy.
As an evaluation of the improvement, the results are compared with parameters taken by conventionally used devices.

\subsection{Spot shape characterization}
\begin{figure}[!htb]
    \centering
    \begin{minipage}[t]{.47\textwidth}
        \centering
          \makebox[\textwidth][c]{\includegraphics[width=1.3\textwidth]{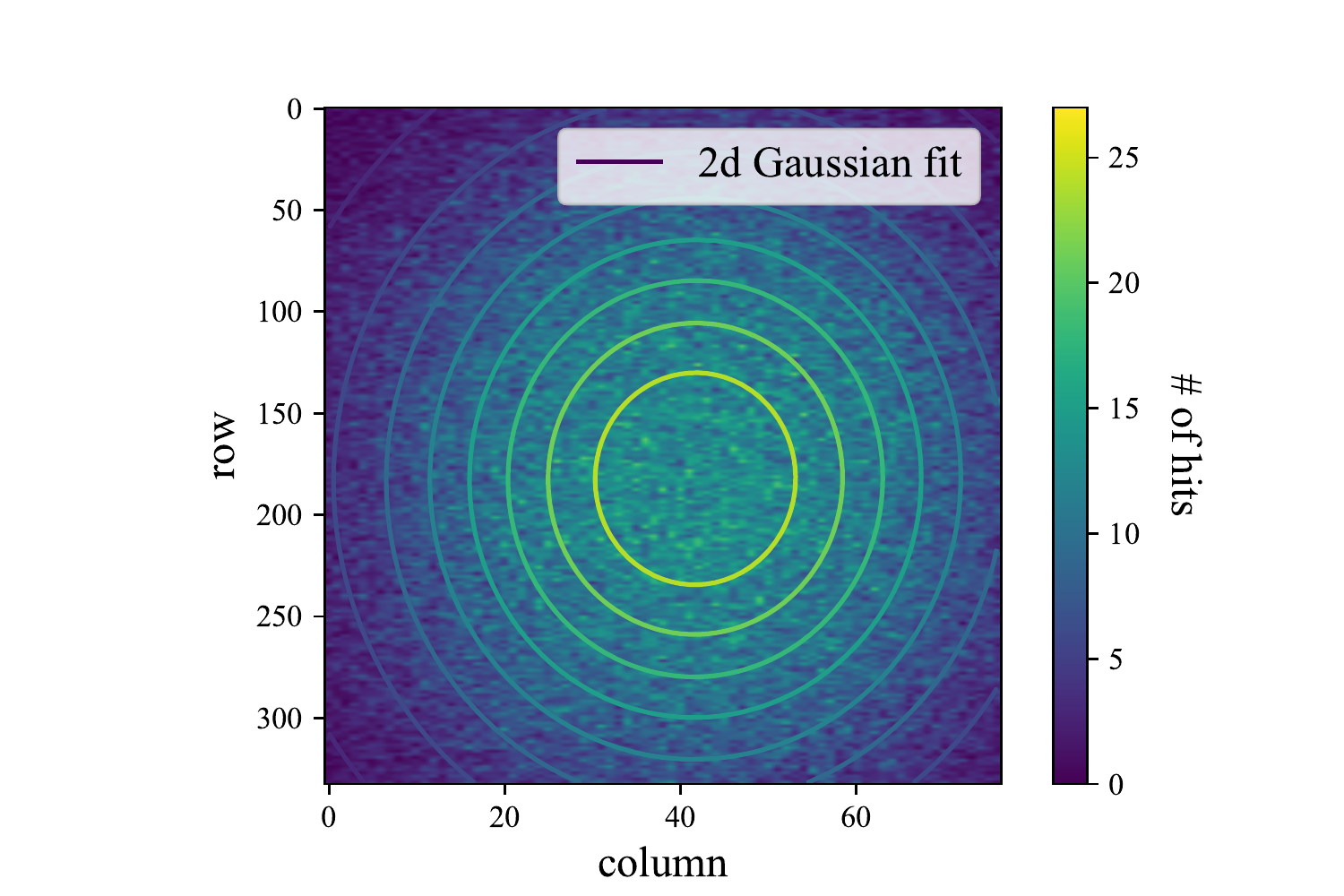}}%
        \caption{Hitmap of a single pencil beam spot. The intensity profile is fitted with a two-dimensional Gaussian function.}
        \label{fig:PBS_shape}
    \end{minipage}%
    \hfill
    \begin{minipage}[t]{.47\textwidth}
        \centering
        \makebox[\textwidth][c]{\includegraphics[width=1.1\textwidth]{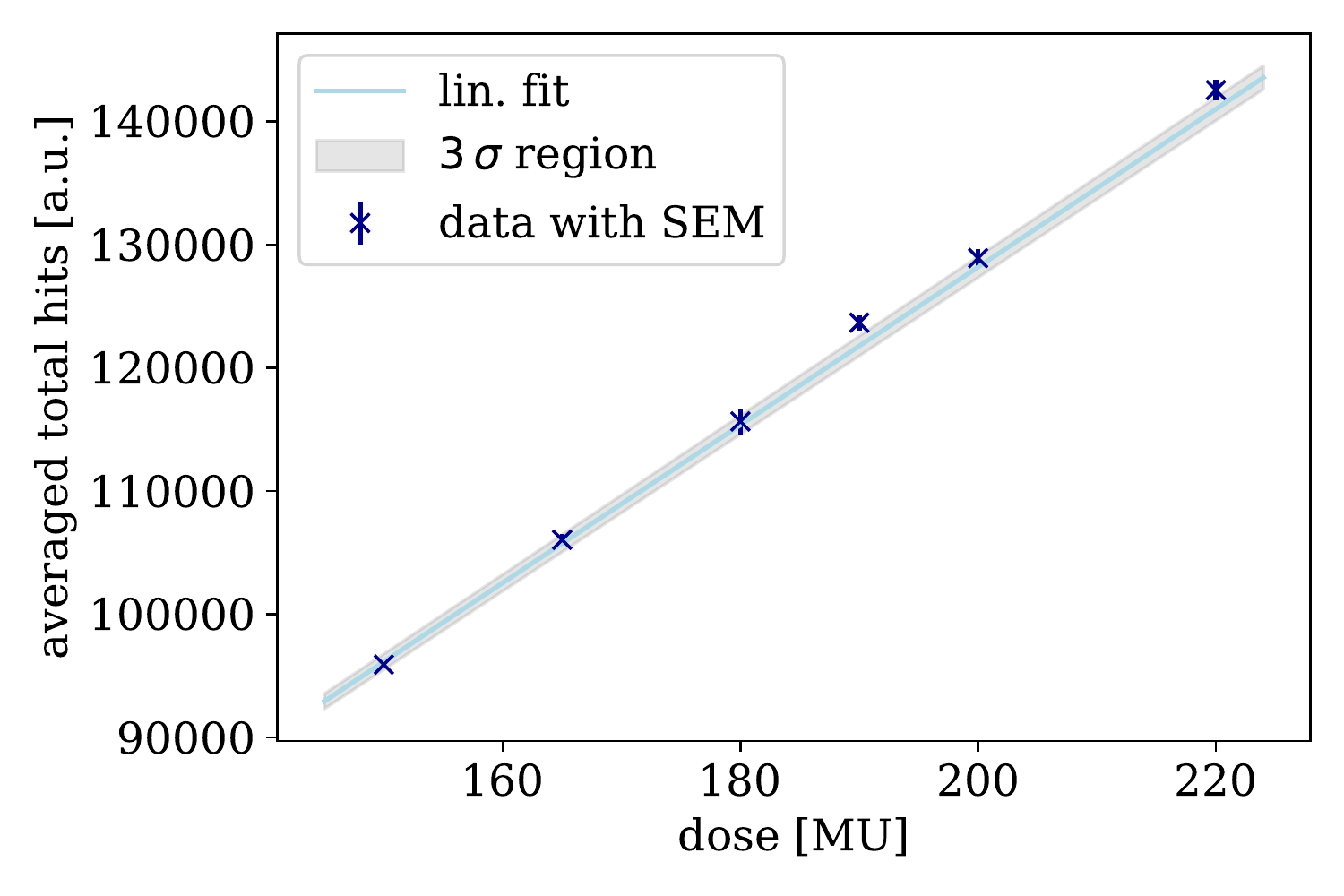}}%
        \caption{Total hits summed across the sensor as a function of the irradiated dose given in facility specified Monitor Units.}
        \label{fig:Dose_calibration}
    \end{minipage}
\end{figure}

The lateral beam profile of a single pencil beam is measured using an external constant frequency trigger not synchronized with the accelerator.
The resulting hitmap is shown in Fig.~\ref{fig:PBS_shape} illustrating the number of hits per pixel across the sensor.

The two-dimensional intensity profile of the pencil beam spot is fitted applying a two-dimensional Gaussian distribution with an accuracy of $\mathcal{X}^2/\rm{DoF} = 1.12$ as indicated by the colored rings, yielding a width of the beam profile of $\sigma = (5.78 \pm 0.03)\,\si{\milli\metre}$. 
The intensity profile agrees well with the measurement using the Lynx PT detector, which yields a spot size of $\sigma_{\rm{Lynx}} = (5.5\pm 0.5)\,\si{\milli\metre}$. The gained uncertainty is one order of magnitude smaller than that of the standard measurement \cite{res_Lynx}.\\
It seems reasonable to design mechanics precise enough to mount the detector on the patient table with an accuracy of $\SI{100}{\micro\meter}$, yielding a precision of the measurement of the beam position well below the $\SI{2}{\milli\meter}$ required in the QA guidelines.

\subsection{Dose constancy verification}
Constancy of the output dose has to be verified with an accuracy of $\SI{3}{\%}$ during the daily quality assurance as recommended by the published QA guidelines for proton therapy \cite{aapm}. 
To use the detector for output constancy testing, the fluence dependent response needs to be calibrated. 
For this purpose, the detector is irradiated several times with the same proton field while varying the dose, given in Monitor Units (MU), integrating the number of hits for any given field over the full detector area. The results of the calibration are shown in Fig.~\ref{fig:Dose_calibration}.
The standard error of the mean number of hits is used to estimate the uncertainty of the dose calibration, leading to an accuracy of the dose measurement of 2.6\%, which is sufficient for daily QA.

\subsection{Range verification}
\begin{figure}[!htb]
    \centering
    \begin{minipage}[t]{.47\textwidth}
        \centering
          \makebox[\textwidth][c]{\includegraphics[width=0.6\textwidth]{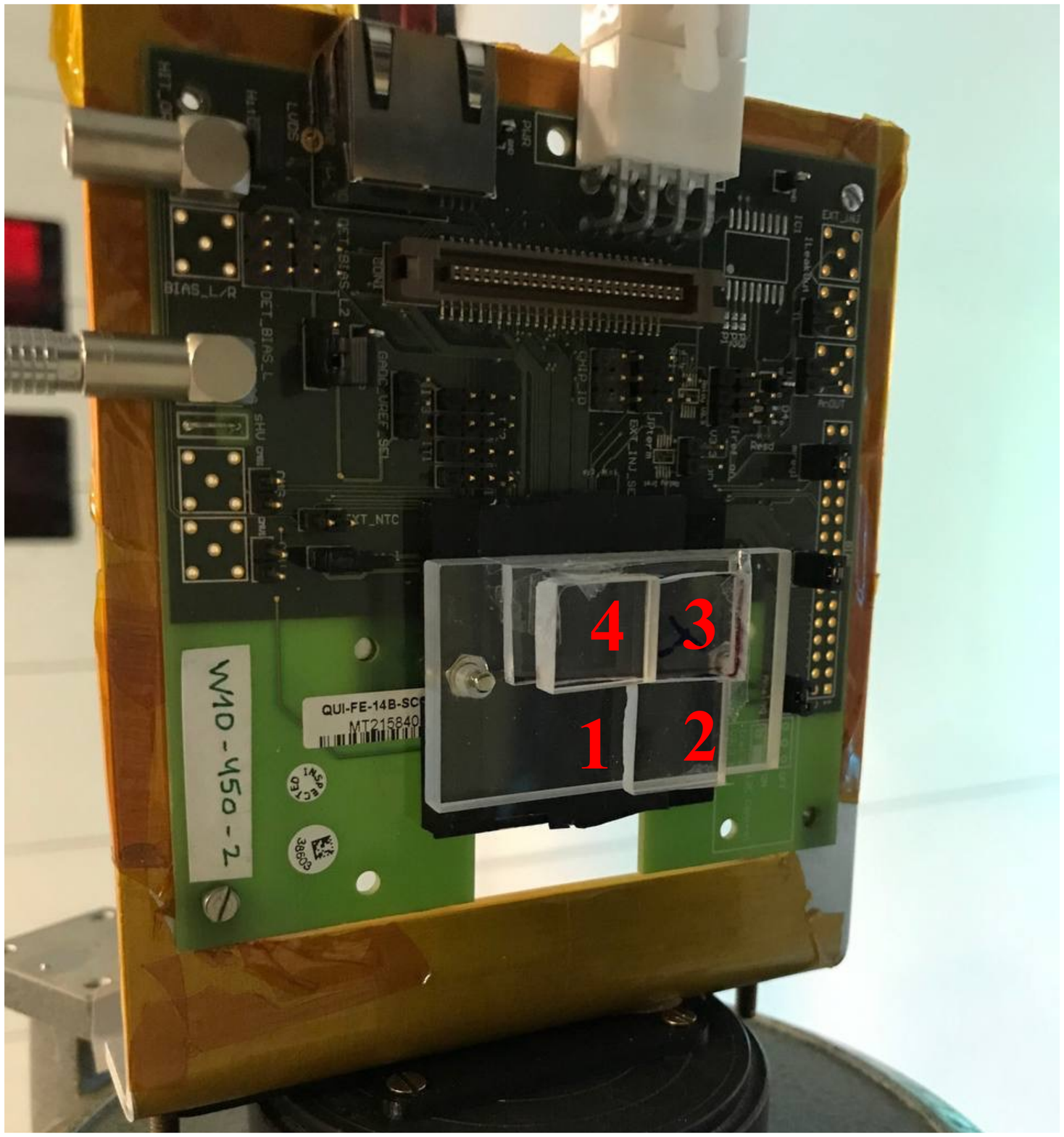}}%
        \caption{Picture of the sensor and the segmented PMMA absorber.}
        \label{fig:Absorber_picture}
    \end{minipage}%
    \hfill
    \begin{minipage}[t]{.47\textwidth}
        \centering
        \makebox[\textwidth][c]{\includegraphics[width=.965\textwidth]{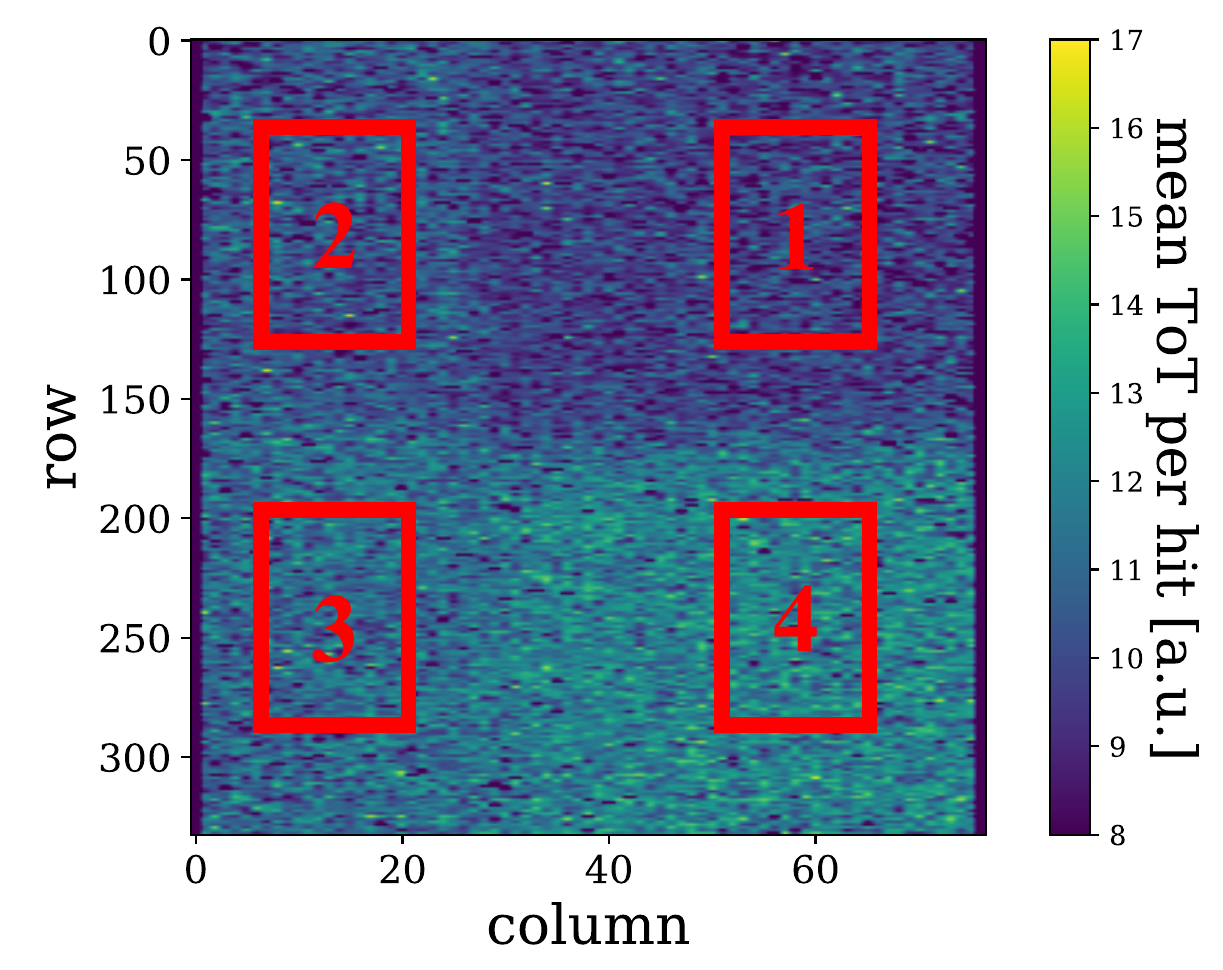}}%
        \caption{Mean ToT map across the sensor, ROIs marked in red.}
        \label{fig:mean_ToT_map}
    \end{minipage}
\end{figure}

Beside the quantities discussed previously, the detector should verify the range of the protons. 
This is done using a PMMA absorber consisting of four regions with thicknesses varying from $\rm{z}_{\rm{region 1}} = \SI{5}{mm}$ up to $\rm{z}_{\rm{region 4}} = \SI{12.5}{mm}$ placed upstream the detector, as shown in Fig.~\ref{fig:Absorber_picture}. 
The difference of the energy deposited in the sensor for different absorber thicknesses is inversely proportional to the energy of the incident protons ($\SI{48}{\mega\eV}$, $\SI{44}{\mega\eV}$, $\SI{40}{\mega\eV}$, $\SI{35}{\mega\eV}$ for $\SI{100}{\MeV}$ incident protons) and thus their range.

A comparison of the energy deposition is done by investigating the resulting mean ToT values for every pixel as shown in Fig.~\ref{fig:mean_ToT_map}. 
The red marks describe the regions of interest (ROI) which are used during the analysis to exclude scattering effects at the edges of the different regions.
An additional absorber with a water equivalent thickness (WET) of $\SI{51}{mm}$ was placed upstream the sensor.
\begin{figure}[htbp]
	\centering
		\includegraphics[width=0.75\textwidth]{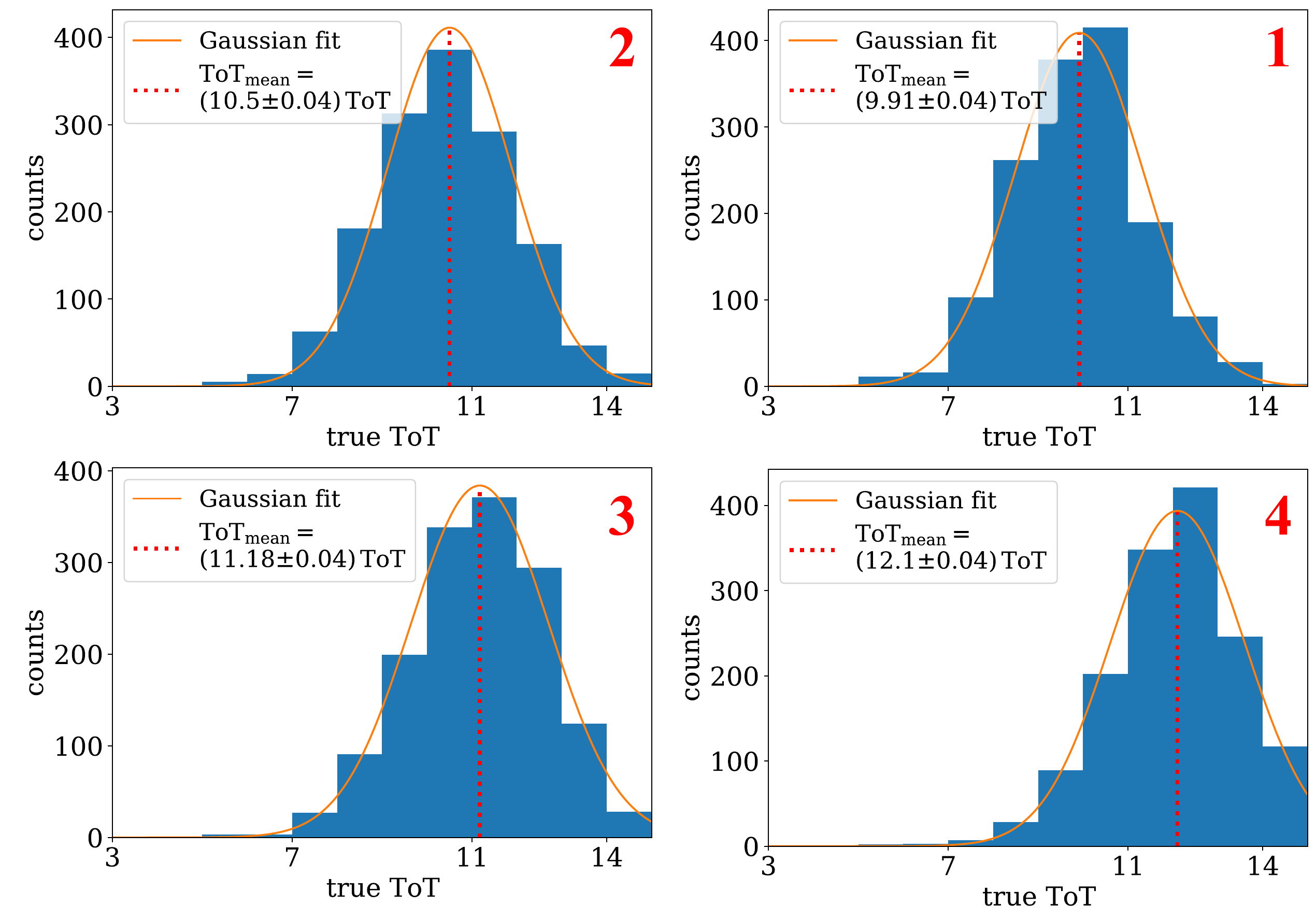}
	\caption{Comparison of the ToT distribution for the different thickness regions of the PMMA absorber.}
	\label{fig:mean_ToT_distribution}
\end{figure}

\begin{figure}[htbp]
	\centering
		\includegraphics[width=0.5\textwidth]{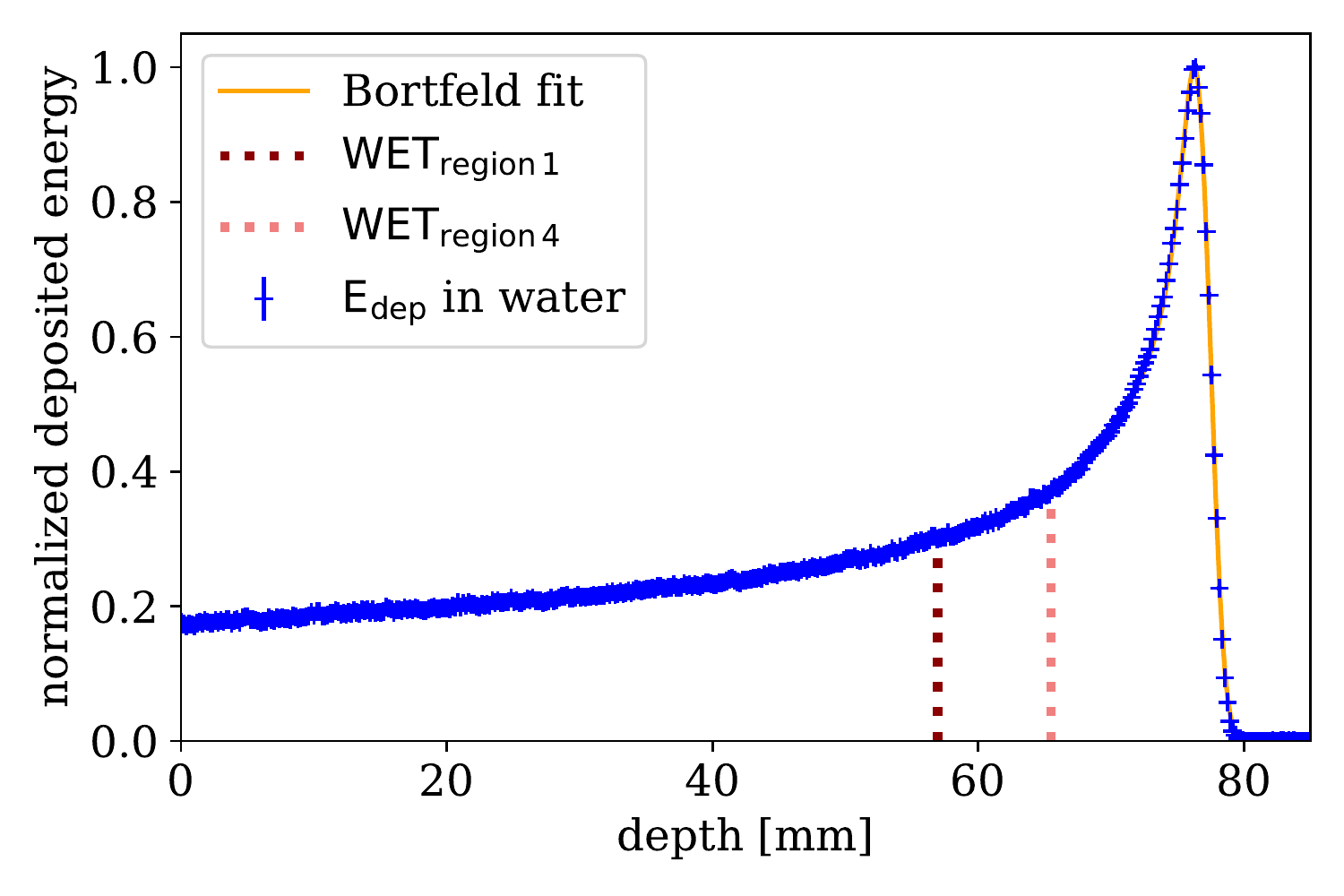}
	\caption{Simulated depth-dose curve in water of $\SI{100}{MeV}$ protons fitted with a Bortfeld function. The WET of the absorbers are indicated.}
	\label{fig:sim_depth_dose_curve}
\end{figure}

To examine those results in more detail, the ToT distribution of every ROI illustrated in Fig.~\ref{fig:mean_ToT_distribution} is fitted with a Gaussian distribution.
The peak of this distribution shifts to higher ToT values with increasing absorber thickness, because protons with lower energy deposit more charge in the sensor.
To monitor proton energy during daily QA we define the ratio of the most likely ToT in different sensor regions, e.g.
\begin{equation}
    \rm{D}_{14} = \frac{\rm{ToT}_{\rm{max, region1}}}{\rm{ToT}_{\rm{max, region4}}}.
\end{equation}
As proof-of-principle for this WET-ratio method, the measured value $\rm{D}_{\rm{14, measured}} = (0.81\,\pm\,0.01$) is compared with the result of a Geant4 (version 10.06.p02) 
simulation of the depth-dose curve of a $\SI{100}{MeV}$ proton beam in water. Shown in
Fig.~\ref{fig:sim_depth_dose_curve} is the deposited energy in $\SI{200}{\micro\m}$ thick slices of water. A Bortfeld function \cite{Bortfeld} is fitted to the data to determine the energy ratio for the same water-equivalent thicknesses as for regions 1 and 4 of the PMMA absorber.
The simulated ratio is $\rm{D}_{\rm{14, simulated}} = 0.80$. 

The uncertainty of this ratio still needs to be evaluated for a final validation of the measurement method. However, the apparent agreement motivates further investigation of it.

\section{Conclusion and perspective}
The application of a high-energy physics detector in daily QA in proton therapy has been presented.
The usage of an ATLAS IBL detector enables the characterization of the beam profile of small spots with one order of magnitude higher spatial resolution compared to commonly used QA devices. 
The results indicate that the requested output constancy with a deviation of $\SI{3}{\%}$ for the daily QA can be verified. First proof-of-principle measurements show promising results of the WET-ratio method for range consistency verification. 
They will be repeated using a precisely machined absorber to determine the detection limit for range variances.

It could be shown that all QA parameters can be measured using an ATLAS pixel detector, which could lead to considerable time savings for the whole procedure. At the same time the spatial resolution of the detector increases the precision of the beam spot size and position measurements, allowing the use of the system for smaller fields and higher dose gradients \cite{delivery}.

\ack
The presented study was supported by the MERCUR-Stiftung graduate school \\ “Präzisionsprotonentherapie – Praxisbezogene Physik und Chemie an der Schnittstelle zur Medizin” (grant number St-2019-0007). Furthermore, the authors would like to thank the IBA PT and the WPE physics team for their support with the irradiations.

\end{document}